\documentclass[twocolumn,floatfix,prb]{revtex4}
\usepackage{graphicx}   
\usepackage{dcolumn}    
\usepackage{bm}         
\def\u2{$\langle u^{2} \rangle$}
\def\lcmo{La$_{0.5}$Ca$_{0.5}$MnO$_{3}$}
\def\mnt{Mn$^{3+}$}
\def\mnf{Mn$^{4+}$}
\def\deg{$^{\circ}$}
\def\dc{$^{\circ}$C}
\def\qmax{Q$_{\mbox{\footnotesize max}}$}
\def\rmin{r$_{\mbox{\footnotesize min}}$}
\def\rmax{r$_{\mbox{\footnotesize max}}$}
\def\inva{\AA$^{-1}$}
\def\tc{T$_{c}$}
\def\tn{T$_{N}$}
\def\oeq{O$_{eq}$}
\def\rwp{R$_{\mbox{\footnotesize wp}}$}
\def\chisq{$\chi^{2}$}
\renewcommand{\sb}[1]{$_{#1}$}
\begin{document}
\title{Neutron diffraction study of average and local structure in \lcmo}
\author {E.~E. Rodriguez}
\author {Th. Proffen} \email{tproffen@lanl.gov}
\author {A. Llobet}
\author {J.~J. Rhyne}

\affiliation{Lujan Neutron Scattering Center, Los Alamos National
Laboratory,Los Alamos, NM.}

\author {J.~F. Mitchell}
\affiliation{Argonne National Laboratory, Argonne, IL}
\date{\today}
%
%
\begin{abstract}
We used neutron powder diffraction to obtain the local and
long-range structure of \lcmo\ at room temperature and 20~K. By
combining Rietveld and pair distribution function analysis of the
total neutron scattering data, we have analyzed the structure of the
compound using two competing models describing the low temperature
phase: first the charge ordered/orbital ordered model and second the
Mn-Mn dimer model. These structural models fit the 20~K neutron
powder diffraction pattern equally well using Rietveld analysis.
Therefore, pair distribution function analysis is used to probe the
local and medium-range structure revealing a system with two
distinctly distorted Mn octahedra and Mn ions with non-integral
valence states. The distorted octahedra differ with the structural
model for the Zener polaron type Mn-Mn dimer picture proposed for
Pr\sb{0.60}Ca\sb{0.40}MnO\sb{3} and order in a similar checkerboard
configuration associated with the CE-type anti-ferromagnetic
structure. Therefore, locally the charge difference and structural
ordering between the two Mn is appreciable enough to describe the
system at 20~K as "partially charge ordered".
\end{abstract}
\pacs{xx}
\maketitle
%
%
\section{Introduction}

Understanding the relation between the magnetic ordering, electronic
ground state, and crystallographic structure in mixed valence
manganites has preoccupied experimentalists and theorists for over
50 years.  Today, manganites continue to be an interesting topic in
solid-state physics because of their high degree of functionality,
strongly correlated electrons, and ability to exhibit fascinating
physical phenomena such as colossal magnetoresistance
(CMR).\cite{covimo99} In 1955 Wollan and Koeler completed their
classic neutron powder diffraction (NPD) study of the effects of
electronic doping on magnetic, structural, and transport properties
of the manganese perovskite
La\sb{1-x}Ca\sb{x}MnO\sb{3}.\cite{woko55} Since LaMnO\sb{3} and
CaMnO\sb{3} have Mn with formal valences of III (d$^{4}$
configuration) and IV (d$^{3}$ configuration) respectively, the
system is hole doped as La$^{3+}$ is substituted by Ca$^{2+}$. The
half-doped compound is an especially interesting composition to
study because it is on a metal to insulator transition interface in
the phase diagram. Moreover, one would expect a ratio of \mnf/\mnt=1
due to an equal amount of divalent (Ca) and trivalent (La)
cations.\cite{woko55}
\par

A qualitative interpretation of Wollan and Koeler's results proposed
by Goodenough predicted that at low temperatures the \mnt\ and \mnf\
ions in the $x=0.5$ compound would order in alternate (110)
planes.\cite{go55} The structure resulting from \mnf/\mnt\ ordering
has been termed the charge-ordered (CO) structure and its associated
magnetic ordering is known as the CE-type anti-ferromagnetic (AFM)
structure. Moreover, the \mnt\ octahedra are Jahn-Teller distorted
and consequently the d$_{z}^{2}$ orbitals order in a zigzag pattern;
this is described as orbital ordering (OO) and is coupled with CO in
Goodenough's model.  Since the \mnf\ octahedra do not distort, their
positions are modulated throughout the structure with an amplitude
that is commensurate with the crystal lattice.
\par

Since the 1990's, Goodenough's interpretation of the charge-ordered
state has been revised and challenged by new experimental studies.
The neutron and x-ray powder diffraction study by Radaelli et
al.\cite{racoma97} suggest that the modulation of the \mnf\
octahedra quasicommensurate at low temperature and still confirms
Goodenough's predictions of charge/orbital ordering
(CO/OO).\cite{racoma97} However, the CO/OO picture has been
challenged by x-ray resonant scattering (RXD), x-ray absorption
near-edge spectroscopy (XANES) and single-crystal neutron
diffraction
studies.\cite{gacobl01,sugabl02,brboan01,daropi02,rodapi02,daropi02b}
These studies conclude that the local structure around the Mn atoms
is incompatible with Goodenough's Mn-O bonding model and that the Mn
atoms do not have integral valence states. In addition, one XANES
study found that one of the Mn octahedra is Jahn-Teller distorted
while the other Mn octahedron lost their local tetrahedral point
group symmetry.\cite{gacobl01} In short, the study concludes that
both octahedra are distorted but in a different manner making the
ionic picture of CO/OO unrealistic.\par

Meanwhile, the study by Daoud-Aladine and Rodriguez-Carvajal
suggests that the CO/OO model is inconsistent with single crystal
diffraction data of Pr$_{0.6}$Ca$_{0.4}$MnO$_{3}$, a system similar
to La$_{1-x}$Ca$_{x}$MnO$_{3}$.\cite{daropi02,rodapi02,daropi02b}
Daoud-Aladine reinterpreted the complex magnetic ordering and the
modulated structural distortion as a consequence of ferromagnetic
(FM) Mn-Mn dimer ordering.  In this model the formal valences of the
Mn are only fractional (Mn$^{+3.5+\delta}$/Mn$^{+3.5-\delta}$) and
all the octahedra are elongated similarly so that two octahedra
connect along their elongated direction.\cite{go55} The FM Mn-Mn
dimer in this model is likened to a Zener polaron.\cite{daropi02} A
Zener polaron is defined as a pair of \mnt\ cations that share a
hole on a bridging oxygen and in which a Zener double exchange
mechanism is responsible for the FM coupling of the Mn
atoms.\cite{zhpa03,zhgo00} It is important to note that orbital
ordering also occurs within the Mn-Mn dimer model. \par

Our study aims to use long-range and short-range analysis of neutron
powder diffraction data to provide further insight into the
modulated distortions seen at lower temperatures.  Rietveld analysis
alone will only yield the average structure of the material. Here we
combine Rietveld analysis with a local structure analysis technique
called the pair distribution function (PDF). The PDF is the Fourier
transform of the total scattering pattern containing both Bragg as
well as diffuse scattering. This approach will result in short,
medium and long range structural information, depending on the
refinement range \rmax. The PDF approach has been used extensively
to study the local structure of other manganites
\cite{bidikw96,prdibi99,bipepr01,eglo02,qiprmi04} and a great
overview of the technique applied to a variety of complex materials
can be found here. \cite{egami}
%
%
\section{Experimental}

Samples were prepared by conventional ceramic methods.  Before
weighing, La\sb{2}O\sb{3} (99.99\%) was fired at 1000~\dc\ in
O\sb{2}, MnO\sb{2} (99.999 \%) was treated in flowing oxygen at
425~\dc\ then slow-cooled; CaCO3 was dried at 100~\dc.
Stoichiometric quantities were mixed and calcined at 900~\dc, then
sintered between 1100~\dc\ and 1380~\dc\ with intermediate grinding
and pelletizing. The final pellet was annealed in O\sb{2} and slow
cooled to ensure stoichiometry.
\par

Neutron powder diffraction data were collected on
NPDF\cite{pregbi02} and HIPD neutron diffractometers located at the
Lujan Neutron Scattering Center at Los Alamos National Laboratory.
The sample of about 10~g was sealed in a cylindrical vanadium tube
with helium exchange gas. Data were collected at temperatures
between T=300~K down to T=20~K in a closed cycle helium
refrigerator. For PDF analysis the data are corrected for detector
deadtime and efficiency, background, absorption, multiple
scattering, inelasticity effects and normalized by the incident flux
and the total sample scattering cross-section to yield the total
scattering structure function, $S(Q)$.  This is Fourier transformed
according to
\begin{equation}
G(r)= \frac{2}{\pi }\int_{0}^{\infty} Q [S(Q)-1] \sin (Qr)\> dQ.
\end{equation}
The data were terminated at a value of \qmax=35~\inva. Data
processing was carried out using the program
PDFgetN.\cite{pegupr00} Data collection and analysis procedures
have been described elsewhere.\cite{egami}

%
%
\section{Results and Analysis}
\subsection{Rietveld Analysis}

\begin{table}[t]
\caption{\label{tab;pbnm_rietveld}RT model used in Rietveld analysis
of 300~K data. Orthorhombic lattice with $Pbnm$ symmetry.  Lattice
constants $a=5.4309(1)$\AA, $b=5.4211(1)$\AA, and $c=7.6400(1)$\AA.
\rwp=4.4\% and \chisq=6.2.}
\begin{ruledtabular}
\begin{tabular}{lcccc}
Atom  & Wyck pos.  &  $x$   &  $y$   &  $z$  \\
\hline
Mn     &  4b   &  0.5       &  0.0        &  0.0       \\
La/Ca  &  4c   & -0.0036(2) &  0.0195(1)  &  0.25      \\
O1     &  4c   &  0.0599(1) &  0.4918(2)  &  0.25      \\
O2     &  8d   &  0.7236(1) &  0.2770(1)  &  0.0312(1) \\
\end{tabular}
\end{ruledtabular}
\end{table}
\begin{figure}[t]
\centering
\includegraphics[angle=0,width=0.9\linewidth]{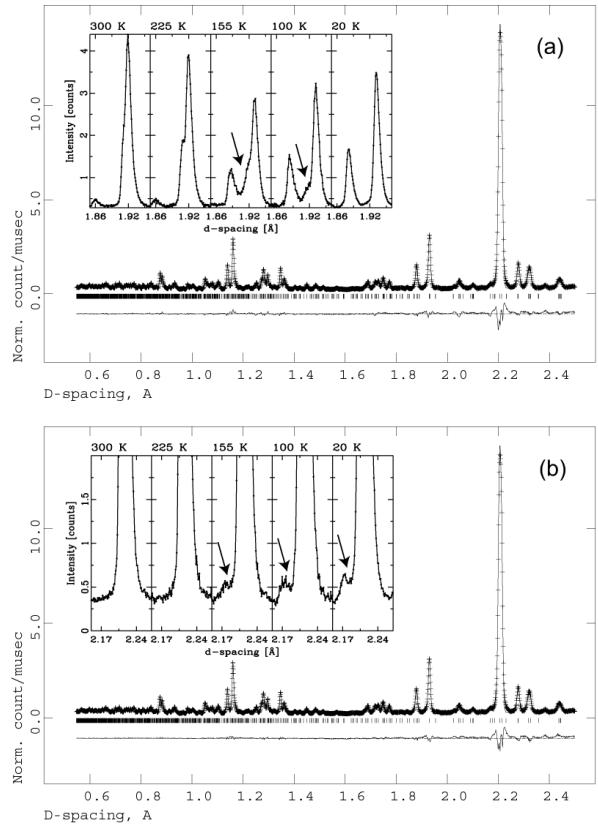}
\caption{(a) NPD pattern at 20~K fitted with the LT-M model.  Arrows
in inset indicate contribution from the room temperature phase. (b)
The LT-O model fit of the 20~K data.  Arrows in inset indicate the
super-lattice peaks.  The bottom line below the tick marks is the
difference between the observed and calculated NPD pattern.  The
temperature evolution in the insets are NPD patterns measured from
HIPD and the fits are done for NPDF data.} \label{fig;riet_p21m}
\end{figure}

\begin{table}[tb]
\caption{\label{tab;p21m_rietveld} LT-M model used in Rietveld
analysis of 20 K data with the values of $\Delta x$, $\Delta x'$,
and the position of atoms refined with $P2_{1}/m$ symmetry
constraints. The monoclinic cell is based on an orthorhombic cell
doubled in the $b$ direction. The $y$ and $z$ coordinates for Mn and
O2 atoms are fixed. $\Delta x=0.0220(1)$, $\Delta x'=0.0103(3)$,
$a=5.47450(7)$\AA, $b=10.88703(13)$\AA, $c=7.51880(8)$\AA,
$\gamma=89.967(3)$\deg. \rwp=5.7\% and \chisq=10.2.}
\begin{ruledtabular}
\begin{tabular}{lcccc}
Atom  & Wyck pos.  &  $x$   &  $y$   &  $z$  \\
\hline
Mn(1) & 2a & 0.0                  &   0.0       &  0.0 \\
Mn(2) & 2b & 0.0                  &   0.5       &  0.0 \\
Mn(3) & 4f & 0.5 - $\Delta x$     &   0.25      &  0.0 \\
La(1) & 2e & 0.4990(5)            &   0.0117(4) &  0.25 \\
La(2) & 2e & 0.4952(5)            &   0.5068(4) &  0.25 \\
La(3) & 2e & 0.0057 - $\Delta x'$ &   0.2601(4) &  0.25 \\
La(4) & 2e & 0.0057 + $\Delta x'$ &   0.7569(3) &  0.25 \\
O1(1) & 2e & -0.0592(6)           &  -0.0133(3) &  0.25 \\
O1(2) & 2e & -0.0538(5)           &   0.4980(3) &  0.25 \\
O1(3) & 2e & 0.5660 - $\Delta x$  &   0.2532(4) &  0.25 \\
O1(4) & 2e & 0.5660 + $\Delta x$  &   0.7470(4) &  0.25 \\
O2(1) & 4f & 0.2248 - $\Delta x$  &   0.1440    &  0.0337  \\
O2(2) & 4f & 0.7198 - $\Delta x$  &   0.1151    & -0.0337 \\
O2(3) & 4f & 0.2248 + $\Delta x$  &   0.6440    &  0.0337 \\
O2(4) & 4f & 0.7198 + $\Delta x$  &   0.6151    & -0.0337 \\
\end{tabular}
\end{ruledtabular}
\end{table}
\begin{table}[tb]
\caption{\label{tab;p21mp_rietveld}Model LT-M' used in Rietveld
analysis of 20~K data.  Atom positions refined with $P2_{1}/m$
symmetry constraints. $a=5.4750(51)$\AA, $b=10.8885(1)$\AA,
$c=7.5194(1)$\AA, $\gamma=89.950(1)$\deg. \rwp=4.5\% and
\chisq=6.5.}
\begin{ruledtabular}
\begin{tabular}{lcccc}
Atom  & Wyck pos.  &  $x$   &  $y$   &  $z$  \\
\hline
Mn(1) & 2a &  0.0       &   0.0       &  0.0 \\
Mn(2) & 2b &  0.0       &   0.5       &  0.0 \\
Mn(3) & 4f &  0.5157(4) &   0.2469(3) &  0.0078(3) \\
La(1) & 2e &  0.0234(5) &   0.2589(3) &  0.25 \\
La(2) & 2e & -0.0090(5) &   0.7644(4) &  0.25 \\
La(3) & 2e &  0.5018(6) &   0.0051(3) &  0.25 \\
La(4) & 2e &  0.4924(5) &   0.5121(4) &  0.25 \\
O1(1) & 2e & -0.0577(6) &   0.0003(4) &  0.25 \\
O1(2) & 2e & -0.0662(5) &   0.4890(3) &  0.25 \\
O1(3) & 2e &  0.5913(4) &   0.2472(2) &  0.25 \\
O1(4) & 2e &  0.5483(5) &   0.7470(4) &  0.25 \\
O2(1) & 4f &  0.2577(4) &   0.1360(3) &  0.0345(4) \\
O2(2) & 4f &  0.2069(4) &   0.6355(3) &  0.0352(4) \\
O2(3) & 4f &  0.7372(3) &   0.1129(4) & -0.0366(4) \\
O2(4) & 4f &  0.7073(4) &   0.6132(2) & -0.0294(4) \\
\end{tabular}
\end{ruledtabular}
\end{table}
\begin{table}[tb]
\caption{\label{tab;p21nm_rietveld} Model LT-O used in Rietveld
analysis of 20~K data.  Atom positions refined with $P2_{1}nm$
symmetry constraints. $a=5.47507(5)$\AA, $b=10.88833(8)$\AA,
$c=7.5193(1)$\AA. \rwp=4.4\% and \chisq=6.2.}
\begin{ruledtabular}
\begin{tabular}{lcccc}
Atom  & Wyck pos.  &  $x$   &  $y$   &  $z$  \\
\hline
Mn(1) & 4b & 0.0176(8)  & 0.3742(4)  & 0.2537(6) \\
Mn(2) & 4b & 0.9923(11) & 0.8764(4)  & 0.2445(6) \\
La(1) & 2a & 0.5017(8)  & 0.3603(3)  & 0.0       \\
La(2) & 2a & 0.5244(8)  & 0.3838(4)  & 0.5       \\
La(3) & 2a & 0.4923(8)  & 0.8676(3)  & 0.0       \\
La(4) & 2a & 0.4906(8)  & 0.8827(3)  & 0.5       \\
O1(1) & 2a & 0.0768(10) & 0.3692(5)  & 0.5       \\
O1(2) & 2a & 0.9448(7)  & 0.3815(4)  & 0.0       \\
O1(3) & 2a & 0.0503(7)  & 0.8811(3)  & 0.5       \\
O1(4) & 2a & 0.9187(9)  & 0.8852(4)  & 0.0       \\
O2(1) & 4b & 0.8020(5)  & 0.5108(2)  & 0.2817(4) \\
O2(2) & 4b & 0.7265(0)  & 0.7605(2)  & 0.2885(4) \\
O2(3) & 4b & 0.2289(10) & 0.7364(3)  & 0.2176(5) \\
O2(4) & 4b & 0.7482(5)  & 0.0104(2)  & 0.2846(3) \\
\end{tabular}
\end{ruledtabular}
\end{table}
\begin{table*}[!bt]
\caption{\label{tab;rietveld_bond} Structural information from
Rietveld analysis of models RT, LT-M, LT-M', and LT-O.  The angles
and lengths are concerned only with the Mn to equatorial O bonds.
Only model RT is used for 300~K data, the rest for 20~K data.}
\begin{ruledtabular}
\begin{tabular}{lcccc}
Model & \multicolumn{2}{c}{Mn-\oeq\ bond distances (\AA)} &
        \multicolumn{2}{c}{Mn-\oeq\ bond angles ($^{\circ}$)} \\
\hline
RT     &  Mn-O      &  2 x 1.94562(1) & Mn-O-Mn        & 161.3861(1)\\
       &  Mn-O      &  2 x 1.94321(1) &                & \\
\hline
LT-M   &  \mnt(1)-O &  2 x 1.93789(6) &\mnt(1)-O-\mnf  & 157.130(1)\\
       &  \mnt(1)-O &  2 x 2.09058(6) &                & 160.1521(9)\\
       &  \mnt(2)-O &  2 x 1.90579(6) &\mnt(2)-O-\mnf  & 161.2382(8)\\
       &  \mnt(2)-O &  2 x 2.08533(6) &                & 162.2580(7)\\
       &  \mnf-O    &  2 x 1.91551(6) &&\\
       &  \mnf-O    &  2 x 1.91464(6) &&\\
\hline
LT-M'  &  \mnt(1)-O &  2 x 1.91209(5) &\mnt(1)-O-\mnf  & 158.7380(7)\\
       &  \mnt(1)-O &  2 x 2.06250(5) &                & 165.4079(4)\\
       &  \mnt(2)-O &  2 x 1.87984(5) &\mnt(2)-O-\mnf  & 159.3062(6)\\
       &  \mnt(2)-O &  2 x 2.03272(5) &                & 162.4260(7)\\
       &  \mnf-O    &  1.87014(5)     &&\\
       &  \mnf-O    &  2.01323(5)     &&\\
       &  \mnf-O    &  1.92520(5)     &&\\
       &  \mnf-O    &  1.95823(5)     &&\\
\hline
LT-O   &  Mn(1)-O   &  1.88581(1)     & Mn(1)-O-Mn(2)  & 157.6742(1)\\
       &  Mn(1)-O   &  1.99869(1)     & Mn(1)-O-Mn(2)  & 162.7741(1)\\
       &  Mn(1)-O   &  1.91047(1)     & Mn(1)-O-Mn(1)  & 161.1720(1)\\
       &  Mn(1)-O   &  2.01610(1)     & Mn(2)-O-Mn(2)  & 163.4689(1)\\
       &  Mn(2)-O   &  1.95475(1)     &&\\
       &  Mn(2)-O   &  2.01104(1)     &&\\
       &  Mn(2)-O   &  1.87888(1)     &&\\
       &  Mn(2)-O   &  2.00133(1)     &&\\
\end{tabular}
\end{ruledtabular}
\end{table*}

The Rietveld refinements were carried out using the GSAS Rietveld
code.\cite{gsas}  At room temperature the structure of \lcmo\ is
single phase and orthorhombic with space group $Pbnm$.  The
parameters of the room temperature phase, denoted RT, are listed in
Table \ref{tab;pbnm_rietveld}. At T=20~K only one phase was observed
and found to have antiferromagnetic ordering of the CE-type.  In
between 225~K and 100~K there is a mixture of the RT and LT phases
as shown in Fig. \ref{fig;riet_p21m}a inset.  As the temperature is
lowered, the low temperature phase develops within the room
temperature phase and its unit cell anisotropically distorts.  This
is similar to other NPD and XRD studies noting coexisting phases
from around 210~K down to low temperatures.\cite{racoma97,hulyer03}
In addition, the room temperature phase develops long-range
ferromagnetism at \tc=195~K and the low temperature phase undergoes
an AFM ordering transition at \tn=155~K. Unlike other NPD studies
our data does not show any contribution from the remnant room
temperature phase at 20~K (Fig. \ref{fig;riet_p21m}a inset). \par

Small satellite peaks observed at 155~K are indicative of the
orbital ordering transition and cannot be fit with model describing
the room temperature phase, denoted model RT. (Fig.
\ref{fig;riet_p21m}b). The most intense satellite peak at
$q=[1,1/2,1]$ is created by the modulated displacement of the Mn
octahedra.  Three structural models were used to fit the satellite
peaks in the 20~K data.  The first model, denoted model LT-M, is the
structural model describing CO/OO at low temperatures as defined by
Radaelli et al.\cite{racoma97} In model LT-M the monoclinic lattice
is doubled in the b-direction and the bond lengths around the \mnf\
are fixed to be approximately 1.915\AA.  Two modulation amplitudes
$\Delta x$ and $\Delta x'$ are applied to the positions of the
equatorial O, \mnf, and half of the La/Ca atoms as shown in Table
\ref{tab;p21m_rietveld}. Along with $\Delta x$ and $\Delta x'$, the
remainder of the atoms' positions are refined with $P2_{1}/m$
symmetry constraints.  The goodness of fit parameters obtained were
a \chisq\ of 10.2 and an \rwp\ of 5.7\%. \par

However, if these constraints are loosened to allow atomic positions
to adjust freely within the symmetry allowed by the space group
$P2_{1}/m$ but with no additional constraints on the \mnf-O bonds,
the \rwp\ and \chisq\ values improve to values of 4.5\% and 6.5
respectively. This model, denoted LT-M', is less constrained than
model LT-M because it does not fix the \mnf-O bond and does not
refine modulation vectors $\Delta x$ and $\Delta x'$. Although the
goodness of fit parameters improve in model LT-M' one should note,
that the number of parameters refined in the non-linear least
squares fit are twice the number as those refined in the model LT-M.
The parameters of model LT-M' are listed in Table
\ref{tab;p21mp_rietveld} and the Mn-O bond lengths are listed in
Table \ref{tab;rietveld_bond}. The LT-M' model has many similarities
with the LT-M model except for the distribution of \mnf-O bond
distances. The resulting bond distances give some indication that
the \mnf\ octahedra may not be as isotropic as assumed in the CO/OO
state. This result anticipates a more rigorous structural analysis
including a short-range order technique to give some more insight.
\par

The third model, denoted LT-O, is the structural model used to
describe the Mn-Mn dimer picture as defined by Daoud-Aladine et al.
\cite{daropi02} In model LT-O the orthorhombic lattice is doubled
along the $b$-axis and the atoms positions are refined with
$P2_{1}nm$ symmetry constraints. Parameters from model LT-O are
shown in Table \ref{tab;p21nm_rietveld} and resulting structural
information in Table \ref{tab;rietveld_bond}.  The third model
yields a \chisq\ of 6.2 and \rwp\ of 4.4\%, which are approximately
equal to those of model LT-M' and slightly better than those of
model LT-M. Again, the number of parameters refined in the least
squares fit is approximately twice the amount in model LT-O compared
to the LT-M model. Tab. \ref{tab;rietveld_bond} lists the bond
lengths and angles of all the models obtained from Rietveld
analysis. \par

\begin{figure}[t]
\centering
\includegraphics[angle=0,width=0.9\linewidth]{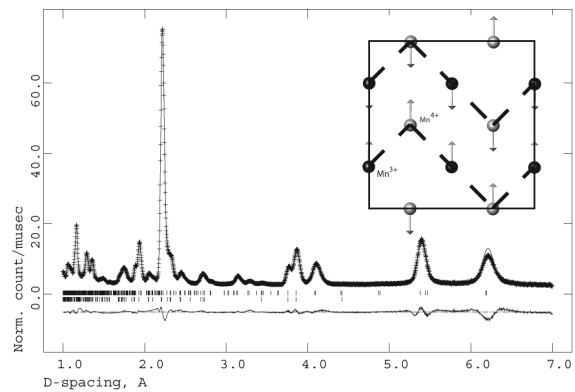}
\caption{NPD pattern of 20~K data measured on HIPD low angle
detector banks.  Upper tick marks indicate contribution from the
CE-type AFM phase and the lower tick marks indicate the nuclear
contribution as defined by model LT-M'. The inset shows the CE-type
anti-ferromagnetic ordering for the \lcmo\ compound at low
temperatures.} \label{fig;riet_afm}
\end{figure}

Finally we have a look at the high d-spacing part of the diffraction
pattern (Fig. \ref{fig;riet_afm}). The magnetic structure was fit by
an AFM phase of the CE-type using a super cell twice the size of the
chemical unit cell determined at 20~K.  The symmetry of the magnetic
phase is consistent with the nuclear phase. As the Shubnikov group
used is $P112_{1}'/m$ which constrains the moments to be in the
$a$-$b$ plane. Four different Mn sites were used to create the
pattern and two different moments were obtained.  The moment value
associated to the so-called \mnt\ site in the CO/OO model is
2.838(11) $\mu_{B}$/Mn and that associated to the \mnf\ is 2.599(10)
$\mu_{B}$/Mn. \par


\subsection{Pair Distribution Function Analysis}

\begin{figure}[b]
\centering
\includegraphics[angle=0,width=0.9\linewidth]{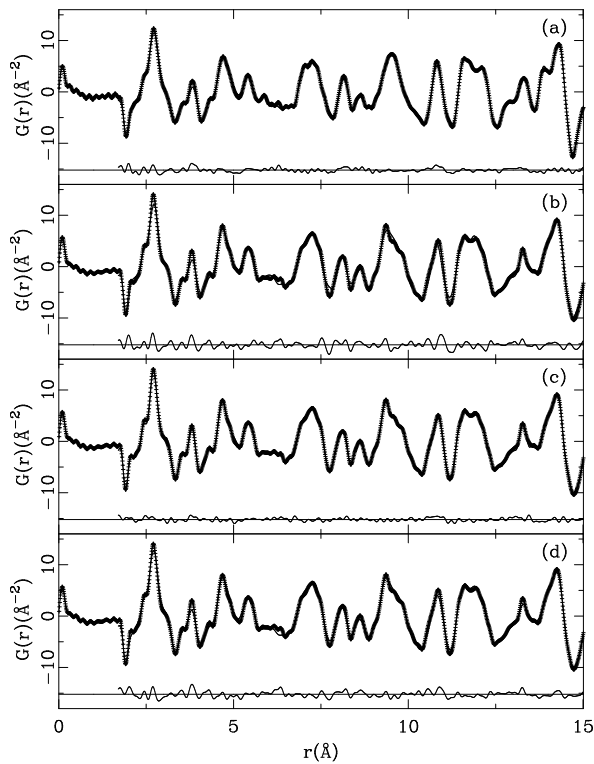}
\caption{PDF refinements over the range \rmin=1.7\AA\ and
\rmax=15\AA: (a) 300~K data using model RT, \rwp=8.8\%; (b) 20~K
data using model RT, \rwp=15.2\%; (c) 20~K data using model LT-M',
\rwp=7.3\% and (d) 20~K data using model LT-O, \rwp=12.8\%.}
\label{fig;pdf_fits15}
\end{figure}
\begin{figure}[b]
\centering
\includegraphics[angle=0,width=0.9\linewidth]{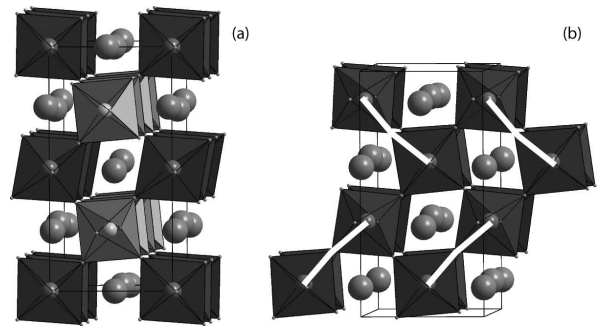}
\caption{ (a)  Resulting structure from the PDF refinement of 20~K
data using model LT-M' for \rmax=15\AA. (b) Resulting structure from
the PDF refinement of 20~K data using model LT-O for \rmax=15\AA.}
\label{fig;pdf_models}
\end{figure}
\begin{table*}[t]
\caption{\label{tab;pdf_bond} Structural information from PDF
analysis of models RT for 300~K data and LT-M' and LT-O for 20~K
data. The angles and lengths are concerned only with the Mn to
equatorial O bonds. \rmax=15\AA\ for all the refinements.}
\begin{ruledtabular}
\begin{tabular}{lcccc}
Model & \multicolumn{2}{c}{Mn-\oeq\ bond distances (\AA)} &
        \multicolumn{2}{c}{Mn-\oeq\ bond angles ($^{\circ}$)} \\
\hline
RT     &  Mn-O      &  2 x 1.942(2)   & Mn-O-Mn        & 161.35(2)\\
       &  Mn-O      &  2 x 1.949(2)   &                & \\
\hline
LT-M'  &  \mnt(1)-O &  2 x 1.932(4)   &\mnt(1)-O-\mnf  & 159.2(3)\\
       &  \mnt(1)-O &  2 x 2.051(4)   &                & 166.7(3)\\
       &  \mnt(2)-O &  2 x 1.917(4)   &\mnt(2)-O-\mnf  & 157.8(2)\\
       &  \mnt(2)-O &  2 x 2.113(4)   &                & 161.2(2)\\
       &  \mnf-O    &      1.948(6)   &&\\
       &  \mnf-O    &      1.845(5)   &&\\
       &  \mnf-O    &      1.969(6)   &&\\
       &  \mnf-O    &      1.884(6)   &&\\
\hline
LT-O   &  Mn(1)-O   &  1.89(4)        & Mn(1)-O-Mn(2)  & 161.1(4)\\
       &  Mn(1)-O   &  1.901(12)      & Mn(1)-O-Mn(2)  & 161.7(5)\\
       &  Mn(1)-O   &  2.018(9)       & Mn(1)-O-Mn(1)  & 158.2(4)\\
       &  Mn(1)-O   &  1.929(10)      & Mn(2)-O-Mn(2)  & 163.5(4)\\
       &  Mn(2)-O   &  2.08(5)        &&\\
       &  Mn(2)-O   &  1.88(3)        &&\\
       &  Mn(2)-O   &  2.01(3)        &&\\
       &  Mn(2)-O   &  1.95(3)        &&\\
\end{tabular}
\end{ruledtabular}
\end{table*}

The results from the Rietveld analysis make it obvious that
analyzing the long-range order in the low temperature phase is
insufficient for deciphering the subtle changes in the Mn octahedra
ordering.  Therefore, we analyzed the information hidden in the
diffuse scattering of the compound through PDF analysis. The PDF can
be calculated from a structural model using the relation
\begin{equation}
G_{c}(r) = \frac{1}{r} \sum_{i}\sum_{j} \left [
           \frac{b_{i}b_{j}}{\langle b \rangle ^{2}}
           \delta (r - r_{ij}) \right ]   \> - 4 \pi r \rho_{0},
\label{eq;igr}
\end{equation}
where the sum goes over all pairs of atoms $i$ and $j$ within the
model crystal separated by $r_{ij}$.  The scattering power of atom
$i$ is $b_{i}$ and $\langle b \rangle$ is the average scattering
power of the sample.To account for the cutoff of $S(Q)$ at
$Q_{max}$, the calculated function $G(r)$ is then convoluted with a
termination function, $\sin(Q_{max}r/r)$. Refinements presented in
this paper were carried out using the program PDFFIT \citep{prbi99}.
The program allows to refine structural parameters such as lattice
parameters, anisotropic atomic displacement parameters, position and
site occupancies. Even though this is similar to the results of a
Rietveld refinement, one needs to realize that the structural model
obtained from PDF analysis is strictly only valid for length scales
corresponding to the $r$-range used for the refinement. This opens
up the possibility to study the local structure on different length
scales by varying the $r$ range refined. In addition to structural
parameters, there are two other corrections to the calculated PDF:
First the finite resolution, $\Delta Q$ of the instrument leads to a
dampening of the PDF intensities by $\exp(-\Delta Q^{2}r^{2}/2)$.
The second correction accounts for changes in the PDF peak width.
The PDF peak width for a pair of atoms $i$ and $j$ is calculated as
\citep{pdffit}
\begin{equation}
  \sigma_{ij} = \sqrt {\sigma_{ij}^{'2} - \frac{\delta}{r_{ij}^{2}} -
                       \frac{\gamma}{r_{ij}} +
                       \alpha^{2} r_{ij}^{2}}.
  \label{eq;sharp}
\end{equation}
The first term $\sigma_{ij}'$ is the PDF peak width of the
structural component due to the atomic displacement parameters. The
next two parameters, $\delta$ and $\gamma$, determine the sharpening
of near neighbor PDF peaks due to correlated motion, in other words
the tendency to move in phase.~\cite{jehegr03,jeprmj99} Finally the
parameter $\alpha$ determines the PDF peak broadening at very high
distances $r$ due to the instrument resolution. Magnetic
correlations are neglected in the PDF analysis because the magnetic
contribution fades quickly with increasing $Q$ compared to the
nuclear contribution. \par

Modeling of the PDF data included two objectives:  First to observe
the differences between the 300~K and 20~K PDF data in order to
determine how the local structure is affected and second to
determine a structural model to describe the 20~K data applying a
strategy similar to the one used in the Rietveld analysis. The PDF
data obtained at T=300~K and T=20~K were fit using RT model over the
range up to \rmax=15\AA\ corresponding to the distance across 3-4
MnO$_{6}$ octahedra. Note that the nearest neighbor peaks in the PDF
are negative because the neutron scattering length of Mn is negative
(see Fig. \ref{fig;pdf_fits15}a and b). One also observed a large
peak at very low values of $r$ in the PDF which is purely due to
noise propagating through the Fourier Transform. As expected, the
data taken at T=300~K are much more accurately described by a
single-cell, orthorhombic lattice than the 20~K data. Since the
lattice parameters and atom positions are refined for both data
sets, the differences seen in Fig. \ref{fig;pdf_fits15}b confirm
that the satellite peaks found in the NPD pattern cannot be modeled
by a single cell and give rise to local structure distortions
affecting the profile of the PDF peaks at 20~K.

After comparing the room temperature and low temperature data, the
low temperature data were refined using two different structural
models, all of which used the doubled cells obtained from Rietveld
analysis as initial structures.  First, we used model LT-M which is
the classic CO/OO structural model constructed with space group
$P2_{1}/m$, in which only the modulation vectors $\Delta x$ and
$\Delta x'$ were refined. Again the refinement range extends to
\rmax=15\AA, yielding a high \rwp\ value of 13.4\% and values of
0.0208(2) and 0.0165(5) for $\Delta x$ and $\Delta x'$ respectively.
The second step of the refinement in this model loosens the previous
modulation vector constraints and allows all the atom positions to
move with $P2_{1}/m$ symmetry constraints. This second step is
similar to the less constrained LT-M' structural model used in the
Rietveld analysis except that the atom positions were refined after
the modulation vectors $\Delta x$ and $\Delta x'$ were applied. The
second set of refinements yields a better \rwp\ value of 7.3\% as
shown in Fig. \ref{fig;pdf_fits15}c and we refer to this model as
LT-M' since we allow the atom positions to move with $P2_{1}/m$
symmetry constraints. \par

The second model is the LT-O model, which allows the atoms to move
in any direction but with $P2_{1}nm$ symmetry constraints. The
result is shown in Fig. \ref{fig;pdf_fits15}d. The \rwp=12.8\%
indicates a much worse agreement of this model compared to the LT-M'
model when refined up to \rmax=15\AA. However, if we extend the
refinement range of the LT-O model to \rmax=45\AA\ (approximately 12
Mn octahedra long diameter), we then find an agreement of
\rwp=12.0\% for the LT-O model which is only slightly worse compared
to \rwp=10.4\% obtained for the LT-M' model. This is consistent with
the Rietveld results yielding a similar agreement for both models.
This means that when a PDF refinement is confined to fit only a
short-range radius of 15\AA, it preferred the LT-M' model to the
LT-O model but for longer correlations, no such preference exists.
The resulting structures from the \rmax\ of 15\AA\ fit for both
models are shown in Fig. \ref{fig;pdf_models}. The LT-M' model still
supports a checkerboard ordering of Mn-O octahedra. Tab.
\ref{tab;pdf_bond} lists the bond lengths and angles of all the
models obtained from PDF analysis for an \rmax=15\AA.\par

\begin{figure*}[t]
\centering
\includegraphics[angle=0,width=0.7\linewidth]{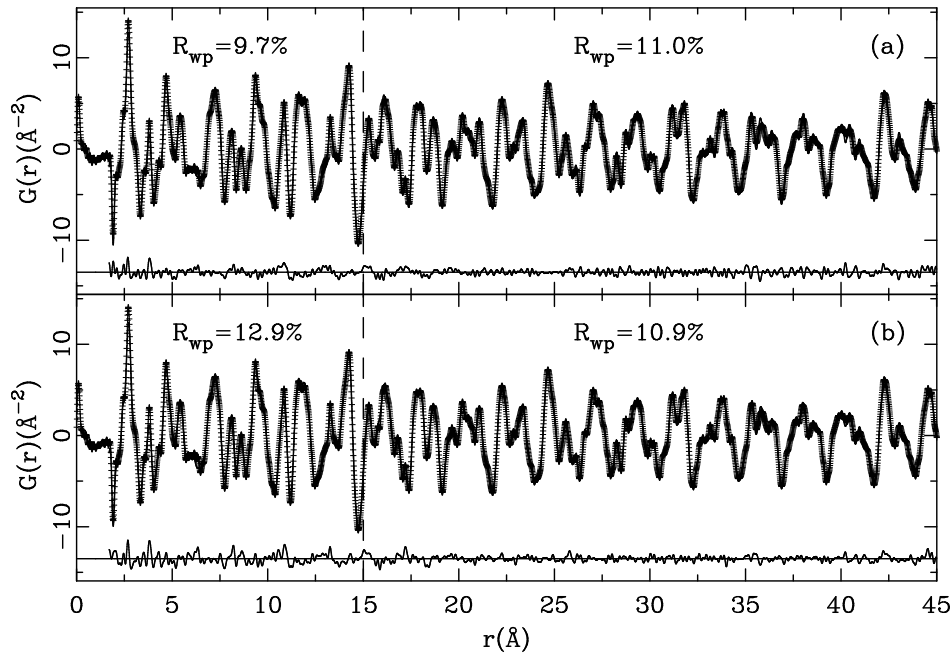}
\caption{PDF refinements over the range \rmin=1.7\AA\ and
\rmax=45\AA: (a) 20~K data using model LT-M', \rwp=10.4\% and (b)
20~K data using model LT-O, \rwp=12.0\%. The \rwp-values indicated
on the panels correspond to the \rwp-values for the corresponding
range in $r$.} \label{fig;pdf_fits45}
\end{figure*}

Based on the refinement of the PDF obtained at T=20~K up to
\rmax=15\AA, model LT-M' gives the best description of the local
structure of \lcmo. However, the results of the Rietveld refinements
gives no such clear distinction between the models. The power of the
PDF technique is the fact that one can probe short as well as medium
range order by selecting the $r$-range used in the
refinement.~\cite{prpa03} In order to investigate the difference
between PDF and Rietveld results, we refined the 20~K PDF data out
to \rmax=45\AA\ using the models LT-M' (Fig. \ref{fig;pdf_fits45}a)
and LT-O (Fig. \ref{fig;pdf_fits45}b). The overall agreement is
\rwp=10.4\% for LT-M' and \rwp=12.0\% for LT-O. However, the
agreement for $15$\AA$<r<45$\AA\ indicated in Fig.
\ref{fig;pdf_fits45} is practically the same. This is consistent
with the findings from the Rietveld refinements and we basically
cannot distinguish between LT-M' and LT-O as model for the medium or
long range average structure of the sample. On the other hand the
local structure ($r<15$\AA) is quite sensitive to the differences of
the two models. From the previous refinements with \rmax=15\AA\ as
well as the \rwp-values shown for $r<15$\AA\ in Fig.
\ref{fig;pdf_fits45} it is obvious that locally the better
structural description is model LT-M' as we have discussed above.
The medium range structure obtained via PDF or the long range
structure obtained via Rietveld refinements on the other hand
supports both models.

%
\section{Discussion}

\begin{table}[b]
\caption{\label{tab;bval} Bond valences calculated from PDF models
LT-M' and LT-O refined to \rmax=15\AA. The estimated standard
deviation on the last digit is given in parenthesis.}
\begin{ruledtabular}
\begin{tabular}{llc}
Model & Mn site & Valence \\
\hline
LT-M'   & \mnt(1) & 3.50(4)+\\
        & \mnt(2) & 3.42(4)+\\
        & \mnf    & 3.99(4)+\\
LT-O    & Mn(1)   & 3.81(4)+\\
        & Mn(2)   & 3.58(4)+\\
\end{tabular}
\end{ruledtabular}
\end{table}

One of the main criticisms about the CO/OO model from X-ray
resonance and absorption studies was that the valence around the Mn
was never found to be fully 4+ or 3+.  Indeed, our study also
concludes the same about the Mn valence state as calculated from
bond valence sums following the relation
\begin{equation}
\label{eq;bval}
V_{i} = \sum_{ij} \exp
        \left [ \frac {r_{ij}^{0} - d_{ij}}{b} \right ]
\end{equation}
where $r$ and $b$ are the so-called bond valence parameters and
$d_{ij}$ is the distance between the central atom $i$ and the
neighboring atom $j$. \cite{bral85,brok91}  The bond valence
calculations for the two models shown in Fig. \ref{fig;pdf_models}
are presented in Tab. \ref{tab;bval}.\par

In the LT-M' model there are 3 Mn sites, where the \mnt(1) and
\mnt(2) sites are Jahn-Teller distorted and the \mnf\ is not. It is
evident that the bond valence of \mnt(1) and \mnt(2) is an
appreciable amount away from the 3+ and both close to 3.5+ using the
bond valence sums. However, the \mnf\ bond valence is actually very
close to 4+.  In the LT-O model obtained from PDF analysis, there
are two different Mn sites, which are all Jahn-Teller distorted and
therefore do not correspond to \mnt\ and \mnf\ atoms. However, even
in this Mn-Mn dimer picture, there are two Mn atoms with different
bond valences as shown in Table \ref{tab;bval} with one close to the
3.5+ value and the other close to 4+. One should keep in mind,
however, that bond valence sum technique has some shortcomings in
describing the bond valence states in the mixed-valence manganites.
\par

\begin{figure}[t]
\centering
\includegraphics[angle=0,width=0.9\linewidth]{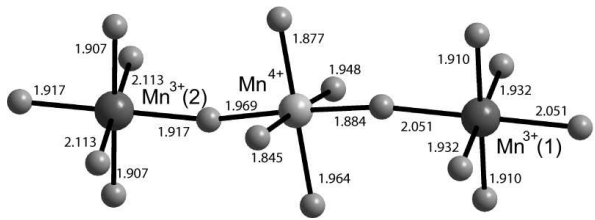}
\caption{Local symmetry around two Mn sites, one corresponding to
the \mnt\ and \mnf\ from the LT-M' model obtained from PDF analysis.
The \mnt\ octahedron is distorted in the expected Jahn-Teller way.
The \mnf\ is not isotropic as expected in the CO/OO picture but off
centered.} \label{fig;mn_oct}
\end{figure}

Another result similar with the local structure analysis from X-ray
data but not incompatible with the CO/OO view is the local symmetry
around the \mnf\ ion in model LT-M' obtained from PDF analysis.  As
demonstrated in other studies, our PDF analysis shows that the local
environment around the \mnf\ atom is anisotropic as shown in Fig.
\ref{fig;mn_oct}. Instead of four equal Mn-O bond lengths on the
equatorial plane of the octahedra, there are two different lengths
in both directions. Therefore, the point group symmetry is lowered
from a tetrahedral point symmetry of $D^{4}_{h}$ down to $C_{s}$,
which only has a mirror plane. This is similar to the XANES study,
where the point group symmetry of the so-called \mnf\ is not
isotropic nor tetrahedral but instead monoclinic.\cite{sugabl02}
This suggests that the \mnf\ is slightly ferroelectric because the
\mnf\ is not centered within the octahedra.  However, no macroscopic
polarity should be detected because the other \mnf\ in the
super-lattice is off-centered in the opposite direction and
therefore should effectively cancel each other out.  This is an
interesting discovery concerning the \mnf\ because it demonstrates
that it can be anisotropic as shown by other short-range structure
studies but in a manner not contradictory to the CO/OO picture.

Another argument against the CO/OO picture is the similar values of
the two Mn moments.  In our Rietveld analysis the moment values are
2.838(11)$\mu_{B}$/Mn and 2.599(10)$\mu_{B}$/Mn for \mnt\ and \mnf\
respectively. Since the moment values are close, this would suggest
that $e_{g}$ electron is localized over two Mn ions as suggested in
the Mn-Mn dimer model. However, other similar studies showing Mn
moments with similar magnitudes conclude that the small difference
in moment values may have to do with incommensurability in the
charge-ordered state, especially in the half-doped
manganites.\cite{racoma97,dlvrji02} One of the studies concludes
that the existence of broadening of certain AFM peaks are a result
of domain boundaries in \mnt\ sublattice, which causes the moment
direction to flip across the magnetic domain
boundary.\cite{racoma97} Therefore, the resulting incommensurability
of the \mnt\ magnetic sublattice causes the refined moment value of
the \mnt\ ion to average out to a lower value.  Indeed, inspection
of the NPD pattern in Fig. \ref{fig;riet_afm} shows that some of the
AFM peaks are broadened more than others and would support the idea
of incommensurability in our charge-ordered structure.\par

%
%
\section{Conclusions}

Our neutron powder diffraction study on \lcmo\ has been analyzed
using Rietveld and Pair distribution function analysis in order to
determine the average and local structure in this controversial
material. Our analysis shows that the ionic CO/OO model and the FM
dimer model succeed to equally model the low temperature charge
ordered state from our powder diffraction data using Rietveld
analysis (long range average structure). Therefore, a local
structural analysis technique was necessary to find the coordination
and bond valence state around the Mn ion as in X-ray local structure
techniques.  Our PDF analysis goes a step further than analyzing the
local environment around the Mn atom (up to 4~\AA). By carrying out
the PDF refinements to \rmax=15\AA\ and to \rmax=45\AA, we can
create a structure that arranges the dissimilarly distorted Mn
octahedra to bridge the gap between local structure analysis and
long-range analysis from Rietveld.
\par

The combined Rietveld/PDF analysis shows that the local structure of
\lcmo\ at 20~K does not support the classic, strictly ionic model
but does not fit the Mn-Mn dimer structure proposed for
Pr\sb{0.6}Ca\sb{0.4}MnO\sb{3} either. Instead, our PDF data are
locally best fit by a model with $P2_{1}/m$ symmetry (LT-M') without
the constraints necessary to create the ionic, integral valence
picture. This structural model supports two different Mn octahedra
that are both distorted but in a different manner.  One shows an
anisotropic distortion associated with the Jahn-Teller effect and a
bond valence state close to 3.5+.  The other has monoclinic
symmetry, is clearly anisotropic, and a bond valence state close to
4+.  The anisotropy in the \mnf\ octahedra implies that the ion is
slightly shifted in one direction within its octahedron. Therefore,
while the system is orbital ordered, it is only partially charge
ordered at low temperatures.  These results reveal the importance of
examining different length scales of a complex system and reveal the
power of pair distribution function when coupled to Rietveld
analysis.

%
%
\begin{acknowledgments}
We thank Ram Seshadri and Nicola Spaldin for helpful discussions and
their insight. This work has benefited from the use of NPDF and HIPD
at the Lujan Center at Los Alamos Neutron Science Center, funded by
DOE Office of Basic Energy Sciences and Los Alamos National
Laboratory funded by Department of Energy under contract
W-7405-ENG-36. The upgrade of NPDF has been funded by NSF through
grant DMR 00-76488.
\end{acknowledgments}
%
%
\bibliography{c:/thomas/documents/bib/thomas/mypub,%
              c:/thomas/documents/bib/diffuse}
\bibliographystyle{apsrev}
%

\end{document}